\documentclass[twocolumn,prfluids,longbibliography,floatfix,a4paper]{revtex4-2}
\usepackage{blindtext}
\usepackage{comment}
\usepackage[english]{babel}
\usepackage[utf8]{inputenc}
\usepackage[T1]{fontenc}
\usepackage{cancel,xcolor}
\usepackage{subcaption}
\usepackage{dsfont}
\usepackage{amsmath}
\usepackage{amssymb}
\usepackage{amsfonts}
\usepackage{amsthm}
\usepackage{graphicx}
\usepackage[scale=0.8]{geometry}
\usepackage{float}

\definecolor{dark-green}{RGB}{0, 128, 0} 

\begin{document}
\title{Gyre Turbulence: Anomalous Dissipation in a Two-Dimensional Ocean Model}
\author{Lennard Miller$^{1,2}$}
\email{lennard.miller@univ-grenoble-alpes.fr}
\author{Bruno Deremble$^{2}$}
\email{bruno.deremble@univ-grenoble-alpes.fr}
\author{Antoine Venaille$^{1}$}
\email{antoine.venaille@ens-lyon.fr}

\affiliation{$^{1}$ENS de Lyon, CNRS, Laboratoire de Physique (UMR CNRS 5672), F-69342 Lyon, France, $^{2}$  Universit\'e Grenoble Alpes, CNRS, INRAE, IRD, Grenoble-INP, Institut des G\'eosciences de l'Environnement, Grenoble, France}

\begin{abstract}
    The exploration of a two-dimensional wind-driven ocean model with no-slip boundaries reveals the existence of a turbulent asymptotic regime where energy dissipation becomes independent of fluid viscosity. This asymptotic flow represents an out-of-equilibrium state, characterized by a vigorous two-dimensional vortex gas superimposed onto a western-intensified gyre. The properties of the vortex gas are elucidated through scaling analysis for detached Prandtl boundary layers, providing a rationalization for the observed anomalous dissipation. The asymptotic regime demonstrates that boundary instabilities alone can be strong enough to evacuate wind-injected energy from the large-scale oceanic circulation.
\end{abstract}

\maketitle

\textit{Introduction.} In three-dimensional turbulence, the transfer of energy from large to small scales results in an energy dissipation rate that remains independent of viscosity, regardless of its smallness \cite{frisch1995turbulence,eyink2006onsager}. This dissipative anomaly is a robust empirical observation sometimes referred to as the zeroth law of turbulence \cite{dubrulle2019beyond}. Conversely, two-dimensional flows are subject to an inverse energy cascade \cite{boffetta2012two}, which results in the self-organization of the flow at the domain scale \cite{bouchet2012statistical}. To ensure efficient dissipation in two-dimensional flows, an additional mechanism is required to generate small-scale structures where dissipation can operate, thus disrupting the inverse cascade. The strong shear near lateral boundaries could serve as a means to create such dissipative structures \cite{deremble2016vorticity}. Numerical studies of the bounded Navier-Stokes equations have examined the relationship between dissipation and viscosity during a dipole-wall collision \cite{clercx2002dissipation, kramer2007vorticity, keetels2011reynolds}. The possibility for dissipation to remain finite in the inviscid limit has been raised \cite{farge2011energy, waidmann2018energy}, yet remains a topic of debate \cite{sutherland2013effect, clercx2017dissipation}.\\ 
The existence of a dissipative anomaly in two-dimensional flows with boundaries would bear significant practical implications, for example in contributing to a deeper understanding of the energy cycle in the ocean \cite{ferrari2009ocean,zhai2010significant,pouquet2013geophysical,pearson2018log}. In fact, classical linear models for the emergence of western intensified currents \cite{vallis2017atmospheric,dalibard2018mathematical}, such as the Gulf Stream or the Kuroshio, do provide a remarkable example of a dissipative anomaly. Here we show that this dissipative anomaly persists in a nonlinear regime, and unveil a new Gyre Turbulence regime with a western intensified mean flow and finite energy dissipation rate. \\

\begin{figure*}
    \includegraphics[width=\textwidth]{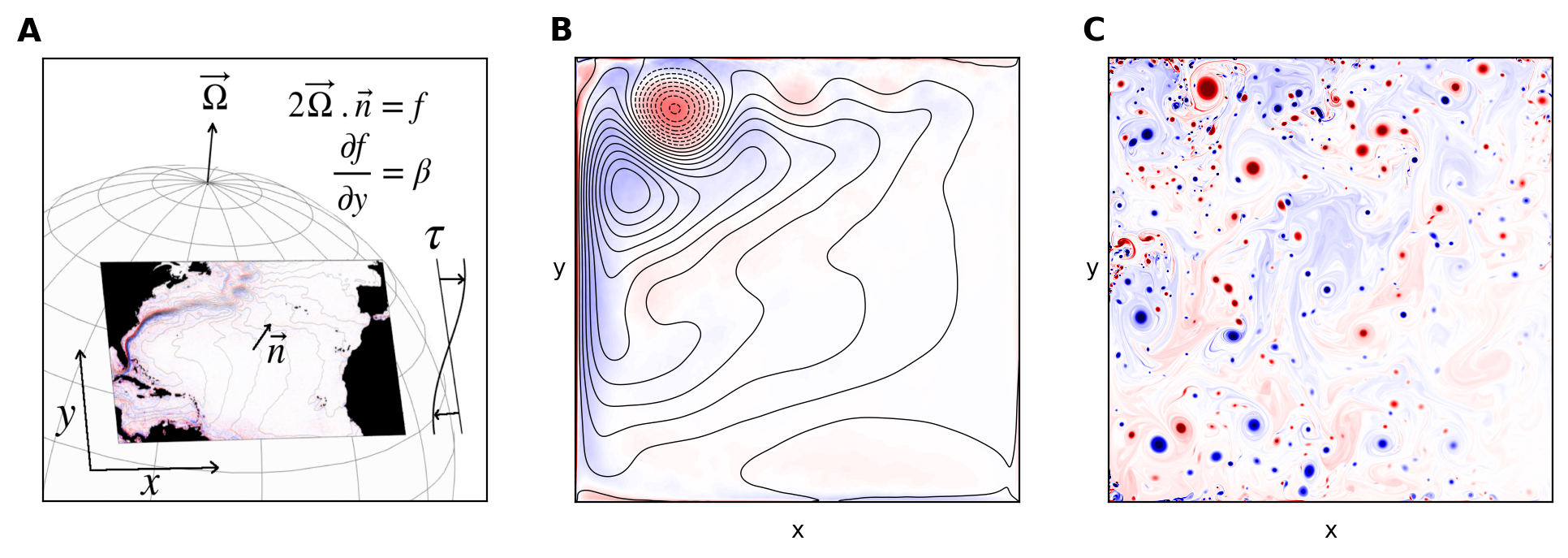}
    \caption{\textbf{A}: The $\beta$-plane model with an inlet for the streamlines  of the time-average sea surface velocity field in the North Atlantic. Colors show the corresponding relative vorticity field. \textbf{B}: Temporal average of the streamlines in the Gyre Turbulence regime ($\nu^* = 2.5\times 10^{-5}$), superimposed on the corresponding relative vorticity field (colors) \textbf{C:} Snapshot of relative vorticity in the Gyre Turbulence regime ($\nu^* = 2\times10^{-6}$). Blue: anticyclonic vortices; red: cyclonic vortices (see Supplementary Material \cite{note:Sup}, movie S2 for an animation).}
    \label{fig: gulf_stream}
\end{figure*}

\textit{Flow model.} The simplest model describing western intensification of oceanic currents is the rigid-lid barotropic quasigeostrophic model on a closed domain tangent to the Earth \cite{vallis2017atmospheric}:
\begin{align}
  \partial_t \omega + J(\psi, \omega) + \beta^* & \partial_x \psi = \nu^* \Delta \omega  -\partial_y\tau , \label{eq:dyn1}\\
    \omega = \Delta \psi, \quad &\tau = - \cos\left(\pi y\right).
    \label{eq:dyn2}
\end{align}
 The term $J(\psi,\omega)=(u\partial_x +v\partial_y) \ \omega$ is the advection of vorticity $\omega$ by the streamfunction $\psi$, with  $u = -\partial_y \psi$ the  zonal ($x$-direction)  and $v = \partial_x \psi$  the meridional (y-direction) velocity components. We solve this two-dimensional model on a square domain with no-slip boundary conditions. Forcing comes from the wind stress curl  $-\partial_y\tau$.  Time and length have been rescaled such that both the length of the domain $L$ and the maximum value of the wind stress $\tau_0$ are $1$. The shape of the forcing corresponds to a single gyre, and is somewhat relevant to the North Atlantic case, with net injection of negative vorticity, as observed in the subtropical gyre (figure~\ref{fig: gulf_stream}A). The only difference to the incompressible two-dimensional Navier-Stokes equations is the term $\beta^* \partial_x \psi$. This term comes from the curl of the Coriolis force, assuming linear variations of the Coriolis parameter in the meridional direction $y$. This framework is called the $beta$-plane approximation, and it captures the effects of differential rotation induced by a rotating planet \cite{vallis2017atmospheric}. 

Earlier studies of the single gyre model with  free-slip boundary conditions exhibited an inertial run-away in the invicid limit: a solution with unrealistically large velocities, and lacking an intensified western boundary current \cite{kamenkovich1995analysis, sheremet1995analysis, ierley1995multiple, sheremet1997eigenanalysis,dupont2004effect,berloff2002material,fox2004wind1}. One way to prevent this inertial run-away is to use no-slip conditions \cite{cessi1990dissipative,nakano2008kuroshio,akuetevi2015dynamics}. It was also noticed that no-slip boundary conditions in two-dimensional turbulence can drastically alter flow organization \cite{clercx1999decaying,clercx2005no,kramer2006beta}.
In the context of a wind-driven shallow water model, \cite{akuetevi2015dynamics} observed that no-slip conditions induced intermittent formation of intense vortices through boundary layer detachment. We show below that this mechanism enables the large-scale flow to remain in an oceanic gyre regime when considering the inviscid limit of (\ref{eq:dyn1} - \ref{eq:dyn2}) (see Supplementary Material \cite{note:Sup}, movie S1 for a direct comparison of the free-slip against the no-slip regimes). \\

\begin{figure*}
\centering
\includegraphics[width = \textwidth]{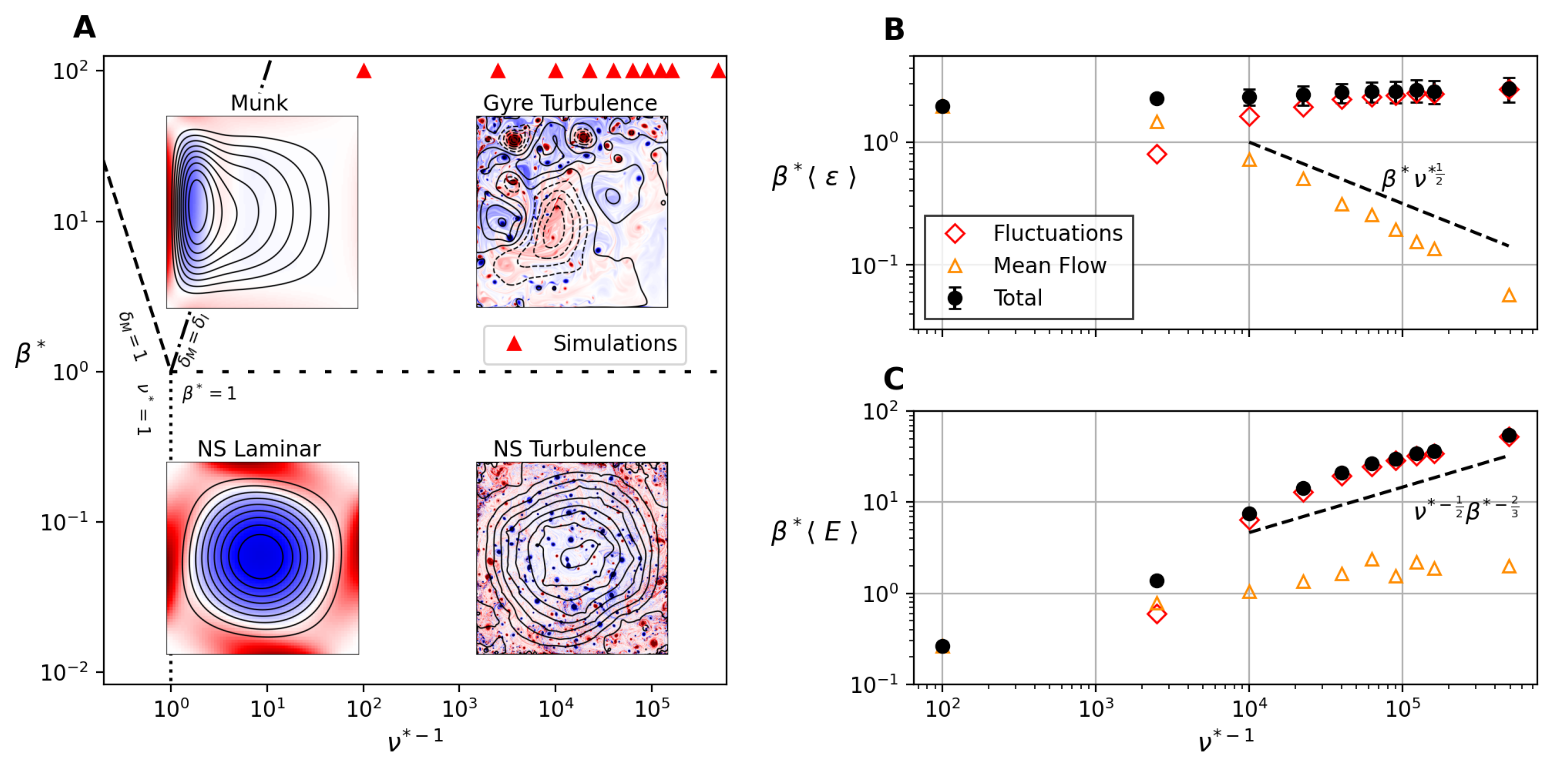}
\caption{\textbf{A}: Parameter space (no-slip boundary conditions). Insets show typical instantaneous vorticity fields (color) and streamfunction (lines) \textbf{B}: Energy dissipation rate, rescaled by the energy injection through a Sverdrup interior. \textbf{C}: Energy rescaled by the energy of the inertial western boundary layer. Errorbars show the standard deviation.}
\label{fig: fin_diss}
\end{figure*}

\textit{Linear dynamics.} Linear theories for wind-driven gyres compute steady states of equations (\ref{eq:dyn1})-(\ref{eq:dyn2}), by neglecting the advection term \cite{vallis2017atmospheric,dalibard2018mathematical}. In the domain bulk, the vorticity equation simplifies into Sverdrup balance, a cornerstone of midlatitude ocean dynamics: $\beta^* v= -\partial_y \tau $, meaning that an injection of negative vorticity is balanced by a southward transport of the fluid. To ensure mass conservation, this interior circulation must be complemented by boundary layers. The majority of this recirculation occurs within the western boundary layer, thereby breaking the East-West symmetry established by the Sverdrup balance. In the viscous solution found by Munk (figure~\ref{fig: fin_diss}A, top left inlet) the boundary layer thickness scales  as $\delta_M = (\nu^*/\beta^*)^{1/3}$ \cite{munk1950wind}, which implies that total dissipation is dominated by contributions from the boundary layer while energy injection comes from the domain bulk. The confinement of energy dissipation in a western boundary layer holds  when viscosity is replaced by other dissipation mechanisms, such as a linear drag \cite{vallis2017atmospheric}. In this scenario, large-scale gyre patterns and therefore energy injection do not depend on details of those linear boundary layers. This is a strong incentive to look for a turbulent dissipative anomaly in this system, adding back nonlinearities into the problem. \\
 
\textit{Parameter regimes.} The linear Munk solution is a limiting case in the $(\nu^*,\beta^*)$ parameter space, corresponding to $\beta^*\gg 1$ and a restricted range of $\nu^*$ to be discussed below. The parameter space can be dissected into four regions (figure~\ref{fig: fin_diss}A): in the limit of weak $\beta^*$, the effects of differential rotation are negligible with respect to other terms and the flow response is equivalent to that of the two-dimensional Navier-Stokes (NS) equations, with a transition from laminar to turbulent flow when $\nu^*$ decreases \cite{clercx2005no}. The same transition from laminar to turbulent flow occurs when differential rotation is important ($\beta^*\gg 1$). When $\nu^* = 10^{-2}$, we observe that the flow consists of a domain interior in which a Sverdrup balance holds with and a stationary western boundary layer, similar to the linear Munk solution. 

When viscosity is decreased, the boundary layer becomes increasingly inertial and its thickness will be governed by $\delta_I =\beta^{*-1}$ \cite{charney1955gulf, vallis2017atmospheric}. The transition between the laminar Munk regime and this inertial regime occurs around $\delta_M = \delta_I$, i.e. $\nu^{*}=1/\beta^{*2}$. When viscosity is decreased further below this threshold the boundary layer becomes unstable, and those instabilities feed the domain with filaments and vortices. As friction is further reduced the system eventually enters into the Gyre Turbulence regime: while the time mean flow remains close to a Sverdrup interior with western intensified boundary layers (figure~\ref{fig: gulf_stream}B), the instantaneous vorticity field is dominated by a vigorous heterogeneous vortex gas which is densest in the north-western corner of the domain (figure~\ref{fig: gulf_stream}C, for an animation see Supplementary Material \cite{note:Sup}, movie S2 ). The gyre structure is not observed on instantaneous streamfunction fields, which is instead dominated by contributions arising from  Rossby basin modes \cite{pedlosky2013geophysical} (not shown here), and to a lesser extent by contributions from vortices.\\

\textit{Energy budget.} The central result of this article is depicted in figure~\ref{fig: fin_diss}B showing that the dissipative anomaly of the linear Munk regime persists in the nonlinear Gyre turbulence regime. Dissipation $\varepsilon$ appears in the energy budget as 
\begin{align}
    \frac{\partial E}{\partial t} = \mathcal{P} -\varepsilon,\quad \mathcal{P}=\int \tau  u \ dA  ,\quad \varepsilon =2\nu^* Z \, ,
    \label{eq:energetics}
\end{align}
where $E=\int (u^2 +v^2)\ dA/2$ is the total energy and $Z=\int \omega^2 \ dA/2$ the total enstrophy. We define time average and fluctuations as 
\begin{align}  \langle\psi\rangle = \lim_{T \to \infty} \frac{1}{T}\int_s^{s + T} \psi \ dt,\quad 
    \psi' = \psi - \langle\psi\rangle,
\end{align}
where the integration is started from a moment $s$ in time after which the system is observed to be in statistical equilibrium, for which $\langle \mathcal{P}\rangle=\langle\varepsilon\rangle$.

In Gyre Turbulence, the average dissipation $\langle \varepsilon \rangle$ is observed to be insensitive to a decrease in $\nu^*$ (Fig.~\ref{fig: fin_diss}B). Our rescaling by $\beta^{*-1}$ shows that the dissipation rate remains close to that predicted by energy injection through an interior flow governed by a Sverdrup balance. However, the dissipation mechanism changes drastically as fluctuations become increasingly important at smaller values of $\nu^*$, while the mean flow contributes only a negligible fraction to the total dissipation. The importance of the fluctuations can also be seen in the total energy of the flow (Fig.~\ref{fig: fin_diss}C). The rescaling by $\beta^{*-1}$ reveals that the total energy of the mean flow remains close to the energy in a western inertial boundary layer, while the magnitude of the fluctuations increases when $\nu^*$ decreases. To rationalize these observations, we describe below the mean flow structure (related to to energy injection), and the production of filaments and vortices (related to energy dissipation).\\ 

\begin{figure}
\centering
\includegraphics[width=\columnwidth]{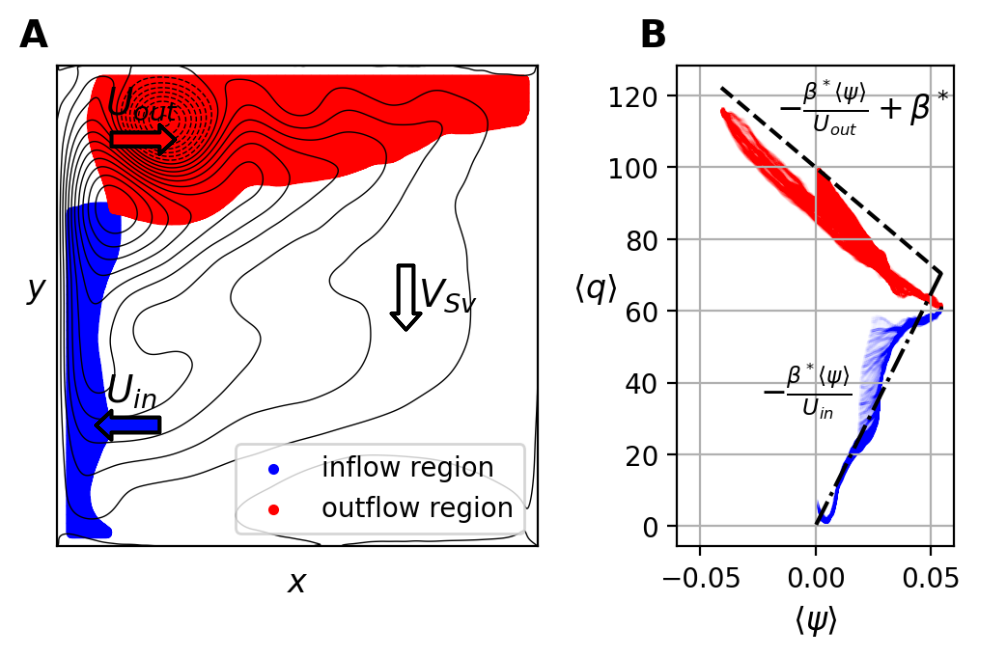}
\caption{\textbf{A}: Contour plot of the mean stream function $\langle \psi \rangle$ at $\nu^* = 2.5\times 10^{-5}$, with colored areas identified as inertial boundary layers. \textbf{B}: $q-\psi$ relations in the inertial boundary layers.}
\label{fig: inertial_theory}
\end{figure}

\textit{Mean flow structure.} In the Gyre Turbulence regime, the time-averaged production term $\langle\mathcal{P}\rangle$ must be independent of $\nu^*$, as it balances the time-averaged dissipation $\langle \epsilon \rangle$. This  constraint, together with the observation in figure~\ref{fig: fin_diss}B that the energy of the mean flow reaches a plateau, suggests that the bulk streamfunction $\langle\psi\rangle$  displayed in figure~\ref{fig: gulf_stream}B only has a weak dependence on $\nu^*$ in the Gyre Turbulence regime. While the order of magnitude of the mean flow agrees with Sverdrup balance in the domain interior (figure~\ref{fig: inertial_theory}A), the gyre pattern is different than the prediction from linear theory, likely due to nonlinear rectification mechanisms in the presence of Rossby waves \cite{fox2004wind2,kramer2006beta}. We observe only weak changes in this pattern when lowering viscosity in the Gyre Turbulence regime (see Supplementary material \cite{note:Sup}, figure 2). 

The gyre pattern is connected to inertial boundary layers with a well defined functional relation between streamfunction and potential vorticity $\langle q\rangle  =\langle \omega\rangle +\beta y$, such that $J(\langle q\rangle,\langle \psi\rangle)=0$. 
We identified two regions where such relations hold (figure~\ref{fig: inertial_theory}), which we will call the inflow (blue) and the outflow (red) layer. 
The inflow layer is consistent with classical theory predicting $\langle q\rangle =-(\beta^*/U_{in}) \langle \psi\rangle$, which leads to a western boundary layer thickness $\delta_I=1/\beta^{*}$, where $U_{in}<0$ is the westward inflow scaling as  $1/\beta^*$ \cite{charney1955gulf, vallis2017atmospheric}. The northward velocity in the inertial boundary layer is $U_I\sim 1$, so that mass transport in this layer compensates the southward Sverdrup bulk transport. No-slip boundary condition is  guaranteed by a Prandtl sublayer with thickness $\delta_P  \sim \sqrt{\nu^*}$. The vorticity within this Prandtl layer is 
\begin{equation}
\omega_{max}=\frac{U_I}{\delta_P}\sim \frac{1} {\sqrt{\nu^*}} . \label{eq:omega_max}
\end{equation}
Assuming that dissipation of the mean flow is governed by these viscous sublayers yields 
\begin{equation}
    \nu^*\int \langle \omega \rangle ^2 dA \sim\nu^* \omega_{max}^2 \delta_P\sim\nu^{*\frac{1}{2}}
\end{equation}
which compares well against the observed dissipation due to the mean flow (see figure~\ref{fig: fin_diss}B).

The outflow layer close to the northern boundary corresponds to a meandering jet with velocity $U_{out}$ and a strong cyclonic recirculation. In this area, we observe a negative correlation between $\langle q \rangle$ and $\langle \psi \rangle$. Stationary  Rossby wave meanders and a stationary vortex on a $beta$-plane with mean flow $U_{out}$ both select the size $\delta_{out} \sim \sqrt{U_{out}/\beta^*}$. We assume that this length sets the vortex and jet width, and that jet transport is set by the transport of the western boundary layer,  which yields  
\begin{equation}
\delta_{out}\sim\beta^{*-2/3} \quad  \textrm{and } \quad U_{out}\sim \beta^{*-1/3}.
\end{equation}
An adaptation of classical Charney theory to this inertial region leads to $\langle q\rangle=-\beta^*/U_{out} \langle \psi\rangle + \beta^*$, which fits well with numerical results (see figure~\ref{fig: inertial_theory}B and Supplementary Material \cite{note:Sup} for further information on boundary layers). \\

\begin{figure*}
\centering
\includegraphics[width = \textwidth]{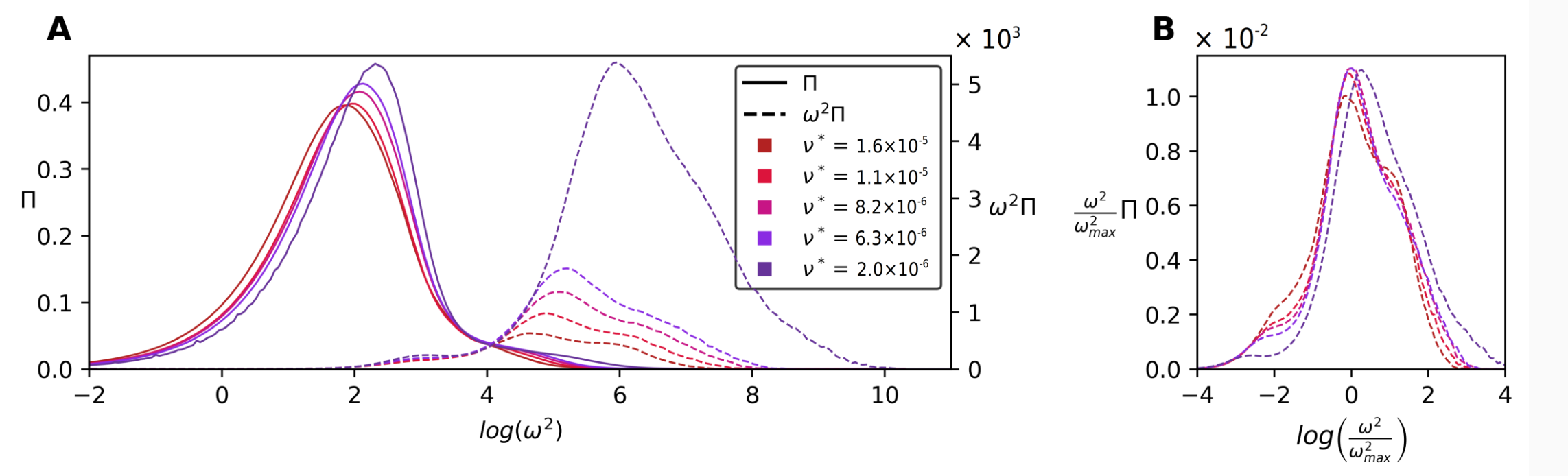}
\caption{\textbf{A}: Probability Distribution functions $\Pi$ of $log(\omega^2)$ for different values of $\nu^*$. Also shown are the values responsible for dissipation, $\omega^{2} \Pi$. \textbf{B}: Values responsible for dissipation after rescaling by $\omega_{max}=1/\sqrt{\nu^*}$. The area under the curves is equal to the total dissipation $\langle \varepsilon \rangle$.}
\label{fig: diss_stat}
\end{figure*}

\textit{Statistics of dissipation.} While the production term $\langle\mathcal{P}\rangle$ depends crucially on the mean flow, we showed that, in the turbulent regime, the dissipation term is dominated by contributions from fluctuations of vorticity (figure~\ref{fig: fin_diss}B). Since dissipation is proportional to enstrophy, we show the probability distribution function of $\omega^2$ in figure~\ref{fig: diss_stat}A (we checked that the distribution of $\omega^{2}$ is close to the distribution of local dissipation $ \vert \nabla \mathbf{u} \vert ^2$).  The core of the distribution $\Pi(\log(\omega^2))$ is close to a Gaussian distribution, similar to recent observations from more comprehensive ocean models \cite{pearson2018log}.  

The peak of $\Pi(\log(\omega^2))$ changes little upon varying the viscosity, revealing that most of the values of vorticity in the bulk are only weakly dependent on $\nu^*$. A much stronger dependence on $\nu^*$ is observed in the tails of the distributions, where we notice important deviations from lognormality. In fact, the average dissipation is dominated by contributions from the tails, which can be seen by plotting the quantity $\omega^2 \Pi$. We show these tails in figure~\ref{fig: diss_stat}B after rescaling the vorticity by $\omega_{max}$ defined in (\ref{eq:omega_max}), collapsing the tails onto a single curve centered around unity. This confirms that the primary mechanism for vorticity injection is the detachment of the Prandtl layer. This scaling for the vorticity can now be used to estimate the energy dissipation rate. \\

\textit{Scaling Analysis of the Vortex Gas.} The cyclonic recirculation we observe in the  mean flow is a signature of the presence of intense cyclonic vortices in this region (see figure~1C). These large vortices result from the coalescence of smaller vortices in the interior. If the vortices grow too large, their drift velocity overcomes the advective velocity from the detached jet, and the vortex collides with the wall, detaching the sublayer over a length $\delta_{out}$ (see Supplementary Material \cite{note:Sup}, movie S2). 

Building upon this observation, the theory for the mean flow and the observed statistics of the dissipation, it is possible to propose a scaling theory that predicts the $\nu^*$-dependency of the energy of the vortex gas. For this, we suppose that the energy is determined by $N$ characteristic vortices of size $\lambda$. If we assume that the vortices are created by the roll-up of the detached sublayer of the western boundary along a length $\delta_{out}$, then the characteristic vortex radius scales as $\lambda \sim \sqrt{\delta_{out} \delta_P}$. Furthermore, in the inviscid limit we observe a balance between the dissipation of these vortices and the injection of energy through the mean Sverdrup flow. The balance between energy dissipation through the characteristic vortices and injection through a Sverdrup interior takes the form
\begin{equation}
    \nu^*N\lambda^2\omega_{max}^2 \sim \beta^{*-1}.
\end{equation}
This determines the area fraction of the vortices as $N\lambda^2 \sim \beta^{*-1}$, which we verify (see Supplementary Material \cite{note:Sup}, figure 2). The total energy of the vortex gas is then given by 
\begin{equation}
    E \sim N\lambda^2(\omega_{max}\lambda)^2 \sim \nu^{*-1/2}\beta^{*-5/3},
\end{equation}
with additional logarithmic corrections \cite{gallet2020vortex}. Although the ideas behind this scaling do not incorporate the fragmentation or coalescence and the resulting inhomogeneity of the vortex gas, a reasonable agreement between simulations and this scaling theory is observed (figure~\ref{fig: fin_diss}C). \\ 

 \textit{Discussion.} Our study challenges the consensus that a weakly dissipated two-dimensional ocean model would lead to energy accumulation due to the absence of a forward energy cascade \cite{vallis2017atmospheric, pouquet2013geophysical}. Instead, no-slip boundaries offer a compelling route to dissipation, sustaining low-energy western-intensified gyres  with a vigorous vortex gas. This Gyre Turbulence regime is a genuine out-of-equilibrium state defying description using  equilibrium statistical mechanics \cite{eyink2006onsager,bouchet2012statistical}, or quasilinear approaches \cite{gallet2013two,frishman2018turbulence}. 
 
Originally discovered in freely decaying 2D flows \cite{mcwilliams1984emergence}, the vortex gas regime emerges in baroclinically unstable stratified quasi-geostrophic flows or any two-dimensional flow with forcing amplifying vorticity extrema \cite{arbic2004baroclinically, thompson2006scaling,venaille2011baroclinic,gallet2020vortex,van2022spontaneous}. Gyre Turbulence, lacking bulk instabilities, generates intense vortex cores solely at boundary layers. The interplay of a western boundary and the $\beta$ effect also seems to prevent crystallisation previously observed on a polar cap \cite{siegelman2022polar}, although this feature may yet occur at lower viscosity not attained in the presented simulations. A quantitative description of the large-scale gyre pattern will require an examination of potential vorticity mixing induced by the vortex gas, as in \cite{gallet2020vortex, gallet2021quantitative,siegelman2023two}. 

In the broader context of the real ocean, turbulent western boundaries have long been recognized as significant energy sinks \cite{zhai2010significant}, with the potential for a linear forward cascade mediated by planetary vorticity gradients \cite{nadiga2010alternating} and intense vortex generation through boundary layer detachment \cite{akuetevi2015dynamics}. Our study reveals that incompressible two-dimensional $\beta$-plane turbulence serves as a minimal model sustaining wind gyres co-existing with mesoscale vortices that contain the majority of the flow energy. However, the two-dimensional model falls short in capturing the fate of these vortices, as evidenced by their energy increase with $\nu^{*-1/2}$. Three-dimensional and ageostrophic effects \cite{pouquet2013geophysical, balwada2022direct}, interactions with bottom topography \cite{dewar2010topographic,nikurashin2013routes,zhang2023spectral}, and air-sea interactions \cite{renault.marchesiello.ea_2019} are expected to play vital roles in dissipating these fine-scale structures. We conjecture that a large part of the energy flux toward these fine scales will continue to be governed by larger scales, a hypothesis currently under examination using more comprehensive ocean models. \\



\textit{Acknowledgements.} This project has received financial support from the CNRS through the 80 Prime  program, and was performed using HPC resources from GENCI-TGCC (Grant 2022-A0130112020). We warmly thank G. Roullet for insightful  inputs on this topic.

\bibliography{biblio}

\begin{thebibliography}{55}%
\makeatletter
\providecommand \@ifxundefined [1]{%
 \@ifx{#1\undefined}
}%
\providecommand \@ifnum [1]{%
 \ifnum #1\expandafter \@firstoftwo
 \else \expandafter \@secondoftwo
 \fi
}%
\providecommand \@ifx [1]{%
 \ifx #1\expandafter \@firstoftwo
 \else \expandafter \@secondoftwo
 \fi
}%
\providecommand \natexlab [1]{#1}%
\providecommand \enquote  [1]{``#1''}%
\providecommand \bibnamefont  [1]{#1}%
\providecommand \bibfnamefont [1]{#1}%
\providecommand \citenamefont [1]{#1}%
\providecommand \href@noop [0]{\@secondoftwo}%
\providecommand \href [0]{\begingroup \@sanitize@url \@href}%
\providecommand \@href[1]{\@@startlink{#1}\@@href}%
\providecommand \@@href[1]{\endgroup#1\@@endlink}%
\providecommand \@sanitize@url [0]{\catcode `\\12\catcode `\$12\catcode
  `\&12\catcode `\#12\catcode `\^12\catcode `\_12\catcode `\%12\relax}%
\providecommand \@@startlink[1]{}%
\providecommand \@@endlink[0]{}%
\providecommand \url  [0]{\begingroup\@sanitize@url \@url }%
\providecommand \@url [1]{\endgroup\@href {#1}{\urlprefix }}%
\providecommand \urlprefix  [0]{URL }%
\providecommand \Eprint [0]{\href }%
\providecommand \doibase [0]{https://doi.org/}%
\providecommand \selectlanguage [0]{\@gobble}%
\providecommand \bibinfo  [0]{\@secondoftwo}%
\providecommand \bibfield  [0]{\@secondoftwo}%
\providecommand \translation [1]{[#1]}%
\providecommand \BibitemOpen [0]{}%
\providecommand \bibitemStop [0]{}%
\providecommand \bibitemNoStop [0]{.\EOS\space}%
\providecommand \EOS [0]{\spacefactor3000\relax}%
\providecommand \BibitemShut  [1]{\csname bibitem#1\endcsname}%
\let\auto@bib@innerbib\@empty
\bibitem [{\citenamefont {Frisch}(1995)}]{frisch1995turbulence}%
  \BibitemOpen
  \bibfield  {author} {\bibinfo {author} {\bibfnamefont {U.}~\bibnamefont
  {Frisch}},\ }\href@noop {} {\emph {\bibinfo {title} {Turbulence: the legacy
  of AN Kolmogorov}}}\ (\bibinfo  {publisher} {Cambridge university press},\
  \bibinfo {year} {1995})\BibitemShut {NoStop}%
\bibitem [{\citenamefont {Eyink}\ and\ \citenamefont
  {Sreenivasan}(2006)}]{eyink2006onsager}%
  \BibitemOpen
  \bibfield  {author} {\bibinfo {author} {\bibfnamefont {G.~L.}\ \bibnamefont
  {Eyink}}\ and\ \bibinfo {author} {\bibfnamefont {K.~R.}\ \bibnamefont
  {Sreenivasan}},\ }\bibfield  {title} {\bibinfo {title} {Onsager and the
  theory of hydrodynamic turbulence},\ }\href@noop {} {\bibfield  {journal}
  {\bibinfo  {journal} {Reviews of modern physics}\ }\textbf {\bibinfo {volume}
  {78}},\ \bibinfo {pages} {87} (\bibinfo {year} {2006})}\BibitemShut {NoStop}%
\bibitem [{\citenamefont {Dubrulle}(2019)}]{dubrulle2019beyond}%
  \BibitemOpen
  \bibfield  {author} {\bibinfo {author} {\bibfnamefont {B.}~\bibnamefont
  {Dubrulle}},\ }\bibfield  {title} {\bibinfo {title} {Beyond kolmogorov
  cascades},\ }\href@noop {} {\bibfield  {journal} {\bibinfo  {journal}
  {Journal of Fluid Mechanics}\ }\textbf {\bibinfo {volume} {867}},\ \bibinfo
  {pages} {P1} (\bibinfo {year} {2019})}\BibitemShut {NoStop}%
\bibitem [{\citenamefont {Boffetta}\ and\ \citenamefont
  {Ecke}(2012)}]{boffetta2012two}%
  \BibitemOpen
  \bibfield  {author} {\bibinfo {author} {\bibfnamefont {G.}~\bibnamefont
  {Boffetta}}\ and\ \bibinfo {author} {\bibfnamefont {R.~E.}\ \bibnamefont
  {Ecke}},\ }\bibfield  {title} {\bibinfo {title} {Two-dimensional
  turbulence},\ }\href@noop {} {\bibfield  {journal} {\bibinfo  {journal}
  {Annual review of fluid mechanics}\ }\textbf {\bibinfo {volume} {44}},\
  \bibinfo {pages} {427} (\bibinfo {year} {2012})}\BibitemShut {NoStop}%
\bibitem [{\citenamefont {Bouchet}\ and\ \citenamefont
  {Venaille}(2012)}]{bouchet2012statistical}%
  \BibitemOpen
  \bibfield  {author} {\bibinfo {author} {\bibfnamefont {F.}~\bibnamefont
  {Bouchet}}\ and\ \bibinfo {author} {\bibfnamefont {A.}~\bibnamefont
  {Venaille}},\ }\bibfield  {title} {\bibinfo {title} {Statistical mechanics of
  two-dimensional and geophysical flows},\ }\href@noop {} {\bibfield  {journal}
  {\bibinfo  {journal} {Physics reports}\ }\textbf {\bibinfo {volume} {515}},\
  \bibinfo {pages} {227} (\bibinfo {year} {2012})}\BibitemShut {NoStop}%
\bibitem [{\citenamefont {Deremble}\ \emph {et~al.}(2016)\citenamefont
  {Deremble}, \citenamefont {Dewar},\ and\ \citenamefont
  {Chassignet}}]{deremble2016vorticity}%
  \BibitemOpen
  \bibfield  {author} {\bibinfo {author} {\bibfnamefont {B.}~\bibnamefont
  {Deremble}}, \bibinfo {author} {\bibfnamefont {W.}~\bibnamefont {Dewar}},\
  and\ \bibinfo {author} {\bibfnamefont {E.}~\bibnamefont {Chassignet}},\
  }\bibfield  {title} {\bibinfo {title} {Vorticity dynamics near sharp
  topographic features},\ }\href@noop {} {\bibfield  {journal} {\bibinfo
  {journal} {J. Mar. Res}\ }\textbf {\bibinfo {volume} {74}},\ \bibinfo {pages}
  {249} (\bibinfo {year} {2016})}\BibitemShut {NoStop}%
\bibitem [{\citenamefont {Clercx}\ and\ \citenamefont {van
  Heijst}(2002)}]{clercx2002dissipation}%
  \BibitemOpen
  \bibfield  {author} {\bibinfo {author} {\bibfnamefont {H.~J.}\ \bibnamefont
  {Clercx}}\ and\ \bibinfo {author} {\bibfnamefont {G.~J.~F.}\ \bibnamefont
  {van Heijst}},\ }\bibfield  {title} {\bibinfo {title} {Dissipation of kinetic
  energy in two-dimensional bounded flows},\ }\href@noop {} {\bibfield
  {journal} {\bibinfo  {journal} {Physical Review E}\ }\textbf {\bibinfo
  {volume} {65}},\ \bibinfo {pages} {066305} (\bibinfo {year}
  {2002})}\BibitemShut {NoStop}%
\bibitem [{\citenamefont {Kramer}\ \emph {et~al.}(2007)\citenamefont {Kramer},
  \citenamefont {Clercx},\ and\ \citenamefont
  {Van~Heijst}}]{kramer2007vorticity}%
  \BibitemOpen
  \bibfield  {author} {\bibinfo {author} {\bibfnamefont {W.}~\bibnamefont
  {Kramer}}, \bibinfo {author} {\bibfnamefont {H.}~\bibnamefont {Clercx}},\
  and\ \bibinfo {author} {\bibfnamefont {G.}~\bibnamefont {Van~Heijst}},\
  }\bibfield  {title} {\bibinfo {title} {Vorticity dynamics of a dipole
  colliding with a no-slip wall},\ }\href@noop {} {\bibfield  {journal}
  {\bibinfo  {journal} {Physics of Fluids}\ }\textbf {\bibinfo {volume} {19}}
  (\bibinfo {year} {2007})}\BibitemShut {NoStop}%
\bibitem [{\citenamefont {Keetels}\ \emph {et~al.}(2011)\citenamefont
  {Keetels}, \citenamefont {Kramer}, \citenamefont {Clercx},\ and\
  \citenamefont {van Heijst}}]{keetels2011reynolds}%
  \BibitemOpen
  \bibfield  {author} {\bibinfo {author} {\bibfnamefont {G.}~\bibnamefont
  {Keetels}}, \bibinfo {author} {\bibfnamefont {W.}~\bibnamefont {Kramer}},
  \bibinfo {author} {\bibfnamefont {H.}~\bibnamefont {Clercx}},\ and\ \bibinfo
  {author} {\bibfnamefont {G.}~\bibnamefont {van Heijst}},\ }\bibfield  {title}
  {\bibinfo {title} {On the reynolds number scaling of vorticity production at
  no-slip walls during vortex-wall collisions},\ }\href@noop {} {\bibfield
  {journal} {\bibinfo  {journal} {Theoretical and computational fluid
  dynamics}\ }\textbf {\bibinfo {volume} {25}},\ \bibinfo {pages} {293}
  (\bibinfo {year} {2011})}\BibitemShut {NoStop}%
\bibitem [{\citenamefont {Nguyen~van yen}\ \emph {et~al.}(2011)\citenamefont
  {Nguyen~van yen}, \citenamefont {Farge},\ and\ \citenamefont
  {Schneider}}]{farge2011energy}%
  \BibitemOpen
  \bibfield  {author} {\bibinfo {author} {\bibfnamefont {R.}~\bibnamefont
  {Nguyen~van yen}}, \bibinfo {author} {\bibfnamefont {M.}~\bibnamefont
  {Farge}},\ and\ \bibinfo {author} {\bibfnamefont {K.}~\bibnamefont
  {Schneider}},\ }\bibfield  {title} {\bibinfo {title} {Energy dissipating
  structures produced by walls in two-dimensional flows at vanishing
  viscosity},\ }\href@noop {} {\bibfield  {journal} {\bibinfo  {journal}
  {Physical Review Letters}\ }\textbf {\bibinfo {volume} {106}},\ \bibinfo
  {pages} {184502} (\bibinfo {year} {2011})}\BibitemShut {NoStop}%
\bibitem [{\citenamefont {Waidmann}\ \emph {et~al.}(2018)\citenamefont
  {Waidmann}, \citenamefont {Klein}, \citenamefont {Farge}, \citenamefont
  {Schneider} \emph {et~al.}}]{waidmann2018energy}%
  \BibitemOpen
  \bibfield  {author} {\bibinfo {author} {\bibfnamefont {M.}~\bibnamefont
  {Waidmann}}, \bibinfo {author} {\bibfnamefont {R.}~\bibnamefont {Klein}},
  \bibinfo {author} {\bibfnamefont {M.}~\bibnamefont {Farge}}, \bibinfo
  {author} {\bibfnamefont {K.}~\bibnamefont {Schneider}}, \emph {et~al.},\
  }\bibfield  {title} {\bibinfo {title} {Energy dissipation caused by boundary
  layer instability at vanishing viscosity},\ }\href@noop {} {\bibfield
  {journal} {\bibinfo  {journal} {Journal of Fluid Mechanics}\ }\textbf
  {\bibinfo {volume} {849}},\ \bibinfo {pages} {676} (\bibinfo {year}
  {2018})}\BibitemShut {NoStop}%
\bibitem [{\citenamefont {Sutherland}\ \emph {et~al.}(2013)\citenamefont
  {Sutherland}, \citenamefont {Macaskill},\ and\ \citenamefont
  {Dritschel}}]{sutherland2013effect}%
  \BibitemOpen
  \bibfield  {author} {\bibinfo {author} {\bibfnamefont {D.}~\bibnamefont
  {Sutherland}}, \bibinfo {author} {\bibfnamefont {C.}~\bibnamefont
  {Macaskill}},\ and\ \bibinfo {author} {\bibfnamefont {D.}~\bibnamefont
  {Dritschel}},\ }\bibfield  {title} {\bibinfo {title} {The effect of slip
  length on vortex rebound from a rigid boundary},\ }\href@noop {} {\bibfield
  {journal} {\bibinfo  {journal} {Physics of Fluids}\ }\textbf {\bibinfo
  {volume} {25}},\ \bibinfo {pages} {093104} (\bibinfo {year}
  {2013})}\BibitemShut {NoStop}%
\bibitem [{\citenamefont {Clercx}\ and\ \citenamefont
  {Van~Heijst}(2017)}]{clercx2017dissipation}%
  \BibitemOpen
  \bibfield  {author} {\bibinfo {author} {\bibfnamefont {H.}~\bibnamefont
  {Clercx}}\ and\ \bibinfo {author} {\bibfnamefont {G.}~\bibnamefont
  {Van~Heijst}},\ }\bibfield  {title} {\bibinfo {title} {Dissipation of
  coherent structures in confined two-dimensional turbulence},\ }\href@noop {}
  {\bibfield  {journal} {\bibinfo  {journal} {Physics of Fluids}\ }\textbf
  {\bibinfo {volume} {29}},\ \bibinfo {pages} {111103} (\bibinfo {year}
  {2017})}\BibitemShut {NoStop}%
\bibitem [{\citenamefont {Ferrari}\ and\ \citenamefont
  {Wunsch}(2009)}]{ferrari2009ocean}%
  \BibitemOpen
  \bibfield  {author} {\bibinfo {author} {\bibfnamefont {R.}~\bibnamefont
  {Ferrari}}\ and\ \bibinfo {author} {\bibfnamefont {C.}~\bibnamefont
  {Wunsch}},\ }\bibfield  {title} {\bibinfo {title} {Ocean circulation kinetic
  energy: Reservoirs, sources, and sinks},\ }\href@noop {} {\bibfield
  {journal} {\bibinfo  {journal} {Annual Review of Fluid Mechanics}\ }\textbf
  {\bibinfo {volume} {41}},\ \bibinfo {pages} {253} (\bibinfo {year}
  {2009})}\BibitemShut {NoStop}%
\bibitem [{\citenamefont {Zhai}\ \emph {et~al.}(2010)\citenamefont {Zhai},
  \citenamefont {Johnson},\ and\ \citenamefont
  {Marshall}}]{zhai2010significant}%
  \BibitemOpen
  \bibfield  {author} {\bibinfo {author} {\bibfnamefont {X.}~\bibnamefont
  {Zhai}}, \bibinfo {author} {\bibfnamefont {H.~L.}\ \bibnamefont {Johnson}},\
  and\ \bibinfo {author} {\bibfnamefont {D.~P.}\ \bibnamefont {Marshall}},\
  }\bibfield  {title} {\bibinfo {title} {Significant sink of ocean-eddy energy
  near western boundaries},\ }\href@noop {} {\bibfield  {journal} {\bibinfo
  {journal} {Nature Geoscience}\ }\textbf {\bibinfo {volume} {3}},\ \bibinfo
  {pages} {608} (\bibinfo {year} {2010})}\BibitemShut {NoStop}%
\bibitem [{\citenamefont {Pouquet}\ and\ \citenamefont
  {Marino}(2013)}]{pouquet2013geophysical}%
  \BibitemOpen
  \bibfield  {author} {\bibinfo {author} {\bibfnamefont {A.}~\bibnamefont
  {Pouquet}}\ and\ \bibinfo {author} {\bibfnamefont {R.}~\bibnamefont
  {Marino}},\ }\bibfield  {title} {\bibinfo {title} {Geophysical turbulence and
  the duality of the energy flow across scales},\ }\href@noop {} {\bibfield
  {journal} {\bibinfo  {journal} {Physical review letters}\ }\textbf {\bibinfo
  {volume} {111}},\ \bibinfo {pages} {234501} (\bibinfo {year}
  {2013})}\BibitemShut {NoStop}%
\bibitem [{\citenamefont {Pearson}\ and\ \citenamefont
  {Fox-Kemper}(2018)}]{pearson2018log}%
  \BibitemOpen
  \bibfield  {author} {\bibinfo {author} {\bibfnamefont {B.}~\bibnamefont
  {Pearson}}\ and\ \bibinfo {author} {\bibfnamefont {B.}~\bibnamefont
  {Fox-Kemper}},\ }\bibfield  {title} {\bibinfo {title} {Log-normal turbulence
  dissipation in global ocean models},\ }\href@noop {} {\bibfield  {journal}
  {\bibinfo  {journal} {Physical review letters}\ }\textbf {\bibinfo {volume}
  {120}},\ \bibinfo {pages} {094501} (\bibinfo {year} {2018})}\BibitemShut
  {NoStop}%
\bibitem [{\citenamefont {Vallis}(2017)}]{vallis2017atmospheric}%
  \BibitemOpen
  \bibfield  {author} {\bibinfo {author} {\bibfnamefont {G.~K.}\ \bibnamefont
  {Vallis}},\ }\href@noop {} {\emph {\bibinfo {title} {Atmospheric and oceanic
  fluid dynamics}}}\ (\bibinfo  {publisher} {Cambridge University Press},\
  \bibinfo {year} {2017})\BibitemShut {NoStop}%
\bibitem [{\citenamefont {Dalibard}\ and\ \citenamefont
  {Saint-Raymond}(2018)}]{dalibard2018mathematical}%
  \BibitemOpen
  \bibfield  {author} {\bibinfo {author} {\bibfnamefont {A.-L.}\ \bibnamefont
  {Dalibard}}\ and\ \bibinfo {author} {\bibfnamefont {L.}~\bibnamefont
  {Saint-Raymond}},\ }\bibfield  {title} {\bibinfo {title} {Mathematical study
  of degenerate boundary layers: A large scale ocean circulation problem},\
  }\href@noop {} {\bibfield  {journal} {\bibinfo  {journal} {American
  Mathematical Society}\ }\textbf {\bibinfo {volume} {253}} (\bibinfo {year}
  {2018})}\BibitemShut {NoStop}%
\bibitem [{\citenamefont {Kamenkovich}\ \emph {et~al.}(1995)\citenamefont
  {Kamenkovich}, \citenamefont {Sheremet}, \citenamefont {Pastushkov},\ and\
  \citenamefont {Belotserkovsky}}]{kamenkovich1995analysis}%
  \BibitemOpen
  \bibfield  {author} {\bibinfo {author} {\bibfnamefont {V.}~\bibnamefont
  {Kamenkovich}}, \bibinfo {author} {\bibfnamefont {V.}~\bibnamefont
  {Sheremet}}, \bibinfo {author} {\bibfnamefont {A.}~\bibnamefont
  {Pastushkov}},\ and\ \bibinfo {author} {\bibfnamefont {S.}~\bibnamefont
  {Belotserkovsky}},\ }\bibfield  {title} {\bibinfo {title} {Analysis of the
  barotropic model of the subtropical gyre in the ocean for finite reynolds
  numbers. part i},\ }\href@noop {} {\bibfield  {journal} {\bibinfo  {journal}
  {Journal of marine research}\ }\textbf {\bibinfo {volume} {53}},\ \bibinfo
  {pages} {959} (\bibinfo {year} {1995})}\BibitemShut {NoStop}%
\bibitem [{\citenamefont {Sheremet}\ \emph {et~al.}(1995)\citenamefont
  {Sheremet}, \citenamefont {Kamenkovich},\ and\ \citenamefont
  {Pastushkov}}]{sheremet1995analysis}%
  \BibitemOpen
  \bibfield  {author} {\bibinfo {author} {\bibfnamefont {V.}~\bibnamefont
  {Sheremet}}, \bibinfo {author} {\bibfnamefont {V.}~\bibnamefont
  {Kamenkovich}},\ and\ \bibinfo {author} {\bibfnamefont {A.}~\bibnamefont
  {Pastushkov}},\ }\bibfield  {title} {\bibinfo {title} {Analysis of the
  barotropic model of the subtropical gyre in the ocean for finite reynolds
  numbers. part ii.},\ }\href@noop {} {\bibfield  {journal} {\bibinfo
  {journal} {Journal of marine research}\ }\textbf {\bibinfo {volume} {53}},\
  \bibinfo {pages} {995} (\bibinfo {year} {1995})}\BibitemShut {NoStop}%
\bibitem [{\citenamefont {Ierley}\ and\ \citenamefont
  {Sheremet}(1995)}]{ierley1995multiple}%
  \BibitemOpen
  \bibfield  {author} {\bibinfo {author} {\bibfnamefont {G.}~\bibnamefont
  {Ierley}}\ and\ \bibinfo {author} {\bibfnamefont {V.}~\bibnamefont
  {Sheremet}},\ }\bibfield  {title} {\bibinfo {title} {Multiple solutions and
  advection-dominated flows in the wind-driven circulation. part i: Slip},\
  }\href@noop {} {\bibfield  {journal} {\bibinfo  {journal} {Journal of marine
  research}\ }\textbf {\bibinfo {volume} {53}},\ \bibinfo {pages} {703}
  (\bibinfo {year} {1995})}\BibitemShut {NoStop}%
\bibitem [{\citenamefont {Sheremet}\ \emph {et~al.}(1997)\citenamefont
  {Sheremet}, \citenamefont {Ierley},\ and\ \citenamefont
  {Kamenkovich}}]{sheremet1997eigenanalysis}%
  \BibitemOpen
  \bibfield  {author} {\bibinfo {author} {\bibfnamefont {V.}~\bibnamefont
  {Sheremet}}, \bibinfo {author} {\bibfnamefont {G.}~\bibnamefont {Ierley}},\
  and\ \bibinfo {author} {\bibfnamefont {V.}~\bibnamefont {Kamenkovich}},\
  }\bibfield  {title} {\bibinfo {title} {Eigenanalysis of the two-dimensional
  wind-driven ocean circulation problem},\ }\href@noop {} {\bibfield  {journal}
  {\bibinfo  {journal} {Journal of marine research}\ }\textbf {\bibinfo
  {volume} {55}},\ \bibinfo {pages} {57} (\bibinfo {year} {1997})}\BibitemShut
  {NoStop}%
\bibitem [{\citenamefont {Dupont}\ and\ \citenamefont
  {Straub}(2004)}]{dupont2004effect}%
  \BibitemOpen
  \bibfield  {author} {\bibinfo {author} {\bibfnamefont {F.}~\bibnamefont
  {Dupont}}\ and\ \bibinfo {author} {\bibfnamefont {D.~N.}\ \bibnamefont
  {Straub}},\ }\bibfield  {title} {\bibinfo {title} {Effect of a wavy wall on
  the single gyre {M}unk problem},\ }\href@noop {} {\bibfield  {journal}
  {\bibinfo  {journal} {Tellus A: Dynamic Meteorology and Oceanography}\
  }\textbf {\bibinfo {volume} {56}},\ \bibinfo {pages} {387} (\bibinfo {year}
  {2004})}\BibitemShut {NoStop}%
\bibitem [{\citenamefont {Berloff}\ \emph {et~al.}(2002)\citenamefont
  {Berloff}, \citenamefont {McWilliams},\ and\ \citenamefont
  {Bracco}}]{berloff2002material}%
  \BibitemOpen
  \bibfield  {author} {\bibinfo {author} {\bibfnamefont {P.~S.}\ \bibnamefont
  {Berloff}}, \bibinfo {author} {\bibfnamefont {J.~C.}\ \bibnamefont
  {McWilliams}},\ and\ \bibinfo {author} {\bibfnamefont {A.}~\bibnamefont
  {Bracco}},\ }\bibfield  {title} {\bibinfo {title} {Material transport in
  oceanic gyres. part i: Phenomenology},\ }\href@noop {} {\bibfield  {journal}
  {\bibinfo  {journal} {Journal of Physical Oceanography}\ }\textbf {\bibinfo
  {volume} {32}},\ \bibinfo {pages} {764} (\bibinfo {year} {2002})}\BibitemShut
  {NoStop}%
\bibitem [{\citenamefont {Fox-Kemper}\ and\ \citenamefont
  {Pedlosky}(2004)}]{fox2004wind1}%
  \BibitemOpen
  \bibfield  {author} {\bibinfo {author} {\bibfnamefont {B.}~\bibnamefont
  {Fox-Kemper}}\ and\ \bibinfo {author} {\bibfnamefont {J.}~\bibnamefont
  {Pedlosky}},\ }\bibfield  {title} {\bibinfo {title} {Wind-driven barotropic
  gyre i: Circulation control by eddy vorticity fluxes to an enhanced removal
  region},\ }\href@noop {} {\bibfield  {journal} {\bibinfo  {journal} {Journal
  of Marine Research}\ }\textbf {\bibinfo {volume} {62}},\ \bibinfo {pages}
  {169} (\bibinfo {year} {2004})}\BibitemShut {NoStop}%
\bibitem [{\citenamefont {Cessi}\ \emph {et~al.}(1990)\citenamefont {Cessi},
  \citenamefont {Condie},\ and\ \citenamefont {Young}}]{cessi1990dissipative}%
  \BibitemOpen
  \bibfield  {author} {\bibinfo {author} {\bibfnamefont {P.}~\bibnamefont
  {Cessi}}, \bibinfo {author} {\bibfnamefont {R.~V.}\ \bibnamefont {Condie}},\
  and\ \bibinfo {author} {\bibfnamefont {W.}~\bibnamefont {Young}},\ }\bibfield
   {title} {\bibinfo {title} {Dissipative dynamics of western boundary
  currents},\ }\href@noop {} {\bibfield  {journal} {\bibinfo  {journal}
  {Journal of Marine Research}\ }\textbf {\bibinfo {volume} {48}},\ \bibinfo
  {pages} {677} (\bibinfo {year} {1990})}\BibitemShut {NoStop}%
\bibitem [{\citenamefont {Nakano}\ \emph {et~al.}(2008)\citenamefont {Nakano},
  \citenamefont {Tsujino},\ and\ \citenamefont {Furue}}]{nakano2008kuroshio}%
  \BibitemOpen
  \bibfield  {author} {\bibinfo {author} {\bibfnamefont {H.}~\bibnamefont
  {Nakano}}, \bibinfo {author} {\bibfnamefont {H.}~\bibnamefont {Tsujino}},\
  and\ \bibinfo {author} {\bibfnamefont {R.}~\bibnamefont {Furue}},\ }\bibfield
   {title} {\bibinfo {title} {The kuroshio current system as a jet and twin
  “relative” recirculation gyres embedded in the sverdrup circulation},\
  }\href@noop {} {\bibfield  {journal} {\bibinfo  {journal} {Dynamics of
  atmospheres and oceans}\ }\textbf {\bibinfo {volume} {45}},\ \bibinfo {pages}
  {135} (\bibinfo {year} {2008})}\BibitemShut {NoStop}%
\bibitem [{\citenamefont {Akuetevi}\ and\ \citenamefont
  {Wirth}(2015)}]{akuetevi2015dynamics}%
  \BibitemOpen
  \bibfield  {author} {\bibinfo {author} {\bibfnamefont {C.~Q.~C.}\
  \bibnamefont {Akuetevi}}\ and\ \bibinfo {author} {\bibfnamefont
  {A.}~\bibnamefont {Wirth}},\ }\bibfield  {title} {\bibinfo {title} {Dynamics
  of turbulent western-boundary currents at low latitude in a shallow-water
  model},\ }\href@noop {} {\bibfield  {journal} {\bibinfo  {journal} {Ocean
  Science}\ }\textbf {\bibinfo {volume} {11}},\ \bibinfo {pages} {471}
  (\bibinfo {year} {2015})}\BibitemShut {NoStop}%
\bibitem [{\citenamefont {Clercx}\ \emph {et~al.}(1999)\citenamefont {Clercx},
  \citenamefont {Maassen},\ and\ \citenamefont
  {Van~Heijst}}]{clercx1999decaying}%
  \BibitemOpen
  \bibfield  {author} {\bibinfo {author} {\bibfnamefont {H.}~\bibnamefont
  {Clercx}}, \bibinfo {author} {\bibfnamefont {S.}~\bibnamefont {Maassen}},\
  and\ \bibinfo {author} {\bibfnamefont {G.}~\bibnamefont {Van~Heijst}},\
  }\bibfield  {title} {\bibinfo {title} {Decaying two-dimensional turbulence in
  square containers with no-slip or stress-free boundaries},\ }\href@noop {}
  {\bibfield  {journal} {\bibinfo  {journal} {Physics of Fluids}\ }\textbf
  {\bibinfo {volume} {11}},\ \bibinfo {pages} {611} (\bibinfo {year}
  {1999})}\BibitemShut {NoStop}%
\bibitem [{\citenamefont {Clercx}\ \emph {et~al.}(2005)\citenamefont {Clercx},
  \citenamefont {Van~Heijst}, \citenamefont {Molenaar},\ and\ \citenamefont
  {Wells}}]{clercx2005no}%
  \BibitemOpen
  \bibfield  {author} {\bibinfo {author} {\bibfnamefont {H.}~\bibnamefont
  {Clercx}}, \bibinfo {author} {\bibfnamefont {G.}~\bibnamefont {Van~Heijst}},
  \bibinfo {author} {\bibfnamefont {D.}~\bibnamefont {Molenaar}},\ and\
  \bibinfo {author} {\bibfnamefont {M.}~\bibnamefont {Wells}},\ }\bibfield
  {title} {\bibinfo {title} {No-slip walls as vorticity sources in
  two-dimensional bounded turbulence},\ }\href@noop {} {\bibfield  {journal}
  {\bibinfo  {journal} {Dynamics of atmospheres and oceans}\ }\textbf {\bibinfo
  {volume} {40}},\ \bibinfo {pages} {3} (\bibinfo {year} {2005})}\BibitemShut
  {NoStop}%
\bibitem [{\citenamefont {Kramer}\ \emph {et~al.}(2006)\citenamefont {Kramer},
  \citenamefont {van Buren}, \citenamefont {Clercx},\ and\ \citenamefont {van
  Heijst}}]{kramer2006beta}%
  \BibitemOpen
  \bibfield  {author} {\bibinfo {author} {\bibfnamefont {W.}~\bibnamefont
  {Kramer}}, \bibinfo {author} {\bibfnamefont {M.}~\bibnamefont {van Buren}},
  \bibinfo {author} {\bibfnamefont {H.}~\bibnamefont {Clercx}},\ and\ \bibinfo
  {author} {\bibfnamefont {G.}~\bibnamefont {van Heijst}},\ }\bibfield  {title}
  {\bibinfo {title} {$\beta$-plane turbulence in a basin with no-slip
  boundaries},\ }\href@noop {} {\bibfield  {journal} {\bibinfo  {journal}
  {Physics of Fluids}\ }\textbf {\bibinfo {volume} {18}},\ \bibinfo {pages}
  {026603} (\bibinfo {year} {2006})}\BibitemShut {NoStop}%
\bibitem [{not()}]{note:Sup}%
  \BibitemOpen
  \href@noop {} {}\bibinfo {note} {See Supplemental Material at [URL will be
  inserted by publisher] for details on numerical methods, boundary layer
  analysis and movies.}\BibitemShut {Stop}%
\bibitem [{\citenamefont {Munk}(1950)}]{munk1950wind}%
  \BibitemOpen
  \bibfield  {author} {\bibinfo {author} {\bibfnamefont {W.~H.}\ \bibnamefont
  {Munk}},\ }\bibfield  {title} {\bibinfo {title} {On the wind-driven ocean
  circulation},\ }\href@noop {} {\bibfield  {journal} {\bibinfo  {journal}
  {Journal of Atmospheric Sciences}\ }\textbf {\bibinfo {volume} {7}},\
  \bibinfo {pages} {80} (\bibinfo {year} {1950})}\BibitemShut {NoStop}%
\bibitem [{\citenamefont {Charney}(1955)}]{charney1955gulf}%
  \BibitemOpen
  \bibfield  {author} {\bibinfo {author} {\bibfnamefont {J.~G.}\ \bibnamefont
  {Charney}},\ }\bibfield  {title} {\bibinfo {title} {The gulf stream as an
  inertial boundary layer},\ }\href@noop {} {\bibfield  {journal} {\bibinfo
  {journal} {Proceedings of the National Academy of Sciences}\ }\textbf
  {\bibinfo {volume} {41}},\ \bibinfo {pages} {731} (\bibinfo {year}
  {1955})}\BibitemShut {NoStop}%
\bibitem [{\citenamefont {Pedlosky}(2013)}]{pedlosky2013geophysical}%
  \BibitemOpen
  \bibfield  {author} {\bibinfo {author} {\bibfnamefont {J.}~\bibnamefont
  {Pedlosky}},\ }\href@noop {} {\emph {\bibinfo {title} {Geophysical fluid
  dynamics}}}\ (\bibinfo  {publisher} {Springer Science \& Business Media},\
  \bibinfo {year} {2013})\BibitemShut {NoStop}%
\bibitem [{\citenamefont {Fox-Kemper}(2004)}]{fox2004wind2}%
  \BibitemOpen
  \bibfield  {author} {\bibinfo {author} {\bibfnamefont {B.}~\bibnamefont
  {Fox-Kemper}},\ }\bibfield  {title} {\bibinfo {title} {Wind-driven barotropic
  gyre ii: Effects of eddies and low interior viscosity},\ }\href@noop {}
  {\bibfield  {journal} {\bibinfo  {journal} {Journal of Marine Research}\
  }\textbf {\bibinfo {volume} {62}},\ \bibinfo {pages} {195} (\bibinfo {year}
  {2004})}\BibitemShut {NoStop}%
\bibitem [{\citenamefont {Gallet}\ and\ \citenamefont
  {Ferrari}(2020)}]{gallet2020vortex}%
  \BibitemOpen
  \bibfield  {author} {\bibinfo {author} {\bibfnamefont {B.}~\bibnamefont
  {Gallet}}\ and\ \bibinfo {author} {\bibfnamefont {R.}~\bibnamefont
  {Ferrari}},\ }\bibfield  {title} {\bibinfo {title} {The vortex gas scaling
  regime of baroclinic turbulence},\ }\href@noop {} {\bibfield  {journal}
  {\bibinfo  {journal} {Proceedings of the National Academy of Sciences}\
  }\textbf {\bibinfo {volume} {117}},\ \bibinfo {pages} {4491} (\bibinfo {year}
  {2020})}\BibitemShut {NoStop}%
\bibitem [{\citenamefont {Gallet}\ and\ \citenamefont
  {Young}(2013)}]{gallet2013two}%
  \BibitemOpen
  \bibfield  {author} {\bibinfo {author} {\bibfnamefont {B.}~\bibnamefont
  {Gallet}}\ and\ \bibinfo {author} {\bibfnamefont {W.}~\bibnamefont {Young}},\
  }\bibfield  {title} {\bibinfo {title} {A two-dimensional vortex condensate at
  high reynolds number},\ }\href@noop {} {\bibfield  {journal} {\bibinfo
  {journal} {Journal of Fluid Mechanics}\ }\textbf {\bibinfo {volume} {715}},\
  \bibinfo {pages} {359} (\bibinfo {year} {2013})}\BibitemShut {NoStop}%
\bibitem [{\citenamefont {Frishman}\ and\ \citenamefont
  {Herbert}(2018)}]{frishman2018turbulence}%
  \BibitemOpen
  \bibfield  {author} {\bibinfo {author} {\bibfnamefont {A.}~\bibnamefont
  {Frishman}}\ and\ \bibinfo {author} {\bibfnamefont {C.}~\bibnamefont
  {Herbert}},\ }\bibfield  {title} {\bibinfo {title} {Turbulence statistics in
  a two-dimensional vortex condensate},\ }\href@noop {} {\bibfield  {journal}
  {\bibinfo  {journal} {Physical review letters}\ }\textbf {\bibinfo {volume}
  {120}},\ \bibinfo {pages} {204505} (\bibinfo {year} {2018})}\BibitemShut
  {NoStop}%
\bibitem [{\citenamefont {McWilliams}(1984)}]{mcwilliams1984emergence}%
  \BibitemOpen
  \bibfield  {author} {\bibinfo {author} {\bibfnamefont {J.~C.}\ \bibnamefont
  {McWilliams}},\ }\bibfield  {title} {\bibinfo {title} {The emergence of
  isolated coherent vortices in turbulent flow},\ }\href@noop {} {\bibfield
  {journal} {\bibinfo  {journal} {Journal of Fluid Mechanics}\ }\textbf
  {\bibinfo {volume} {146}},\ \bibinfo {pages} {21} (\bibinfo {year}
  {1984})}\BibitemShut {NoStop}%
\bibitem [{\citenamefont {Arbic}\ and\ \citenamefont
  {Flierl}(2004)}]{arbic2004baroclinically}%
  \BibitemOpen
  \bibfield  {author} {\bibinfo {author} {\bibfnamefont {B.~K.}\ \bibnamefont
  {Arbic}}\ and\ \bibinfo {author} {\bibfnamefont {G.~R.}\ \bibnamefont
  {Flierl}},\ }\bibfield  {title} {\bibinfo {title} {Baroclinically unstable
  geostrophic turbulence in the limits of strong and weak bottom ekman
  friction: Application to midocean eddies},\ }\href@noop {} {\bibfield
  {journal} {\bibinfo  {journal} {Journal of Physical Oceanography}\ }\textbf
  {\bibinfo {volume} {34}},\ \bibinfo {pages} {2257} (\bibinfo {year}
  {2004})}\BibitemShut {NoStop}%
\bibitem [{\citenamefont {Thompson}\ and\ \citenamefont
  {Young}(2006)}]{thompson2006scaling}%
  \BibitemOpen
  \bibfield  {author} {\bibinfo {author} {\bibfnamefont {A.~F.}\ \bibnamefont
  {Thompson}}\ and\ \bibinfo {author} {\bibfnamefont {W.~R.}\ \bibnamefont
  {Young}},\ }\bibfield  {title} {\bibinfo {title} {Scaling baroclinic eddy
  fluxes: Vortices and energy balance},\ }\href@noop {} {\bibfield  {journal}
  {\bibinfo  {journal} {Journal of physical oceanography}\ }\textbf {\bibinfo
  {volume} {36}},\ \bibinfo {pages} {720} (\bibinfo {year} {2006})}\BibitemShut
  {NoStop}%
\bibitem [{\citenamefont {Venaille}\ \emph {et~al.}(2011)\citenamefont
  {Venaille}, \citenamefont {Vallis},\ and\ \citenamefont
  {Smith}}]{venaille2011baroclinic}%
  \BibitemOpen
  \bibfield  {author} {\bibinfo {author} {\bibfnamefont {A.}~\bibnamefont
  {Venaille}}, \bibinfo {author} {\bibfnamefont {G.~K.}\ \bibnamefont
  {Vallis}},\ and\ \bibinfo {author} {\bibfnamefont {K.~S.}\ \bibnamefont
  {Smith}},\ }\bibfield  {title} {\bibinfo {title} {Baroclinic turbulence in
  the ocean: Analysis with primitive equation and quasigeostrophic
  simulations},\ }\href@noop {} {\bibfield  {journal} {\bibinfo  {journal}
  {Journal of Physical Oceanography}\ }\textbf {\bibinfo {volume} {41}},\
  \bibinfo {pages} {1605} (\bibinfo {year} {2011})}\BibitemShut {NoStop}%
\bibitem [{\citenamefont {van Kan}\ \emph {et~al.}(2022)\citenamefont {van
  Kan}, \citenamefont {Favier}, \citenamefont {Julien},\ and\ \citenamefont
  {Knobloch}}]{van2022spontaneous}%
  \BibitemOpen
  \bibfield  {author} {\bibinfo {author} {\bibfnamefont {A.}~\bibnamefont {van
  Kan}}, \bibinfo {author} {\bibfnamefont {B.}~\bibnamefont {Favier}}, \bibinfo
  {author} {\bibfnamefont {K.}~\bibnamefont {Julien}},\ and\ \bibinfo {author}
  {\bibfnamefont {E.}~\bibnamefont {Knobloch}},\ }\bibfield  {title} {\bibinfo
  {title} {Spontaneous suppression of inverse energy cascade in
  instability-driven 2-d turbulence},\ }\href@noop {} {\bibfield  {journal}
  {\bibinfo  {journal} {Journal of Fluid Mechanics}\ }\textbf {\bibinfo
  {volume} {952}},\ \bibinfo {pages} {R4} (\bibinfo {year} {2022})}\BibitemShut
  {NoStop}%
\bibitem [{\citenamefont {Siegelman}\ \emph {et~al.}(2022)\citenamefont
  {Siegelman}, \citenamefont {Young},\ and\ \citenamefont
  {Ingersoll}}]{siegelman2022polar}%
  \BibitemOpen
  \bibfield  {author} {\bibinfo {author} {\bibfnamefont {L.}~\bibnamefont
  {Siegelman}}, \bibinfo {author} {\bibfnamefont {W.~R.}\ \bibnamefont
  {Young}},\ and\ \bibinfo {author} {\bibfnamefont {A.~P.}\ \bibnamefont
  {Ingersoll}},\ }\bibfield  {title} {\bibinfo {title} {Polar vortex crystals:
  Emergence and structure},\ }\href@noop {} {\bibfield  {journal} {\bibinfo
  {journal} {Proceedings of the National Academy of Sciences}\ }\textbf
  {\bibinfo {volume} {119}},\ \bibinfo {pages} {e2120486119} (\bibinfo {year}
  {2022})}\BibitemShut {NoStop}%
\bibitem [{\citenamefont {Gallet}\ and\ \citenamefont
  {Ferrari}(2021)}]{gallet2021quantitative}%
  \BibitemOpen
  \bibfield  {author} {\bibinfo {author} {\bibfnamefont {B.}~\bibnamefont
  {Gallet}}\ and\ \bibinfo {author} {\bibfnamefont {R.}~\bibnamefont
  {Ferrari}},\ }\bibfield  {title} {\bibinfo {title} {A quantitative scaling
  theory for meridional heat transport in planetary atmospheres and oceans},\
  }\href@noop {} {\bibfield  {journal} {\bibinfo  {journal} {AGU Advances}\
  }\textbf {\bibinfo {volume} {2}},\ \bibinfo {pages} {e2020AV000362} (\bibinfo
  {year} {2021})}\BibitemShut {NoStop}%
\bibitem [{\citenamefont {Siegelman}\ and\ \citenamefont
  {Young}(2023)}]{siegelman2023two}%
  \BibitemOpen
  \bibfield  {author} {\bibinfo {author} {\bibfnamefont {L.}~\bibnamefont
  {Siegelman}}\ and\ \bibinfo {author} {\bibfnamefont {W.}~\bibnamefont
  {Young}},\ }\bibfield  {title} {\bibinfo {title} {Two-dimensional turbulence
  above topography: Vortices and potential vorticity homogenization},\
  }\href@noop {} {\bibfield  {journal} {\bibinfo  {journal} {Proceedings of the
  National Academy of Sciences}\ }\textbf {\bibinfo {volume} {120}},\ \bibinfo
  {pages} {e2308018120} (\bibinfo {year} {2023})}\BibitemShut {NoStop}%
\bibitem [{\citenamefont {Nadiga}\ and\ \citenamefont
  {Straub}(2010)}]{nadiga2010alternating}%
  \BibitemOpen
  \bibfield  {author} {\bibinfo {author} {\bibfnamefont {B.}~\bibnamefont
  {Nadiga}}\ and\ \bibinfo {author} {\bibfnamefont {D.}~\bibnamefont
  {Straub}},\ }\bibfield  {title} {\bibinfo {title} {Alternating zonal jets and
  energy fluxes in barotropic wind-driven gyres},\ }\href@noop {} {\bibfield
  {journal} {\bibinfo  {journal} {Ocean Modelling}\ }\textbf {\bibinfo {volume}
  {33}},\ \bibinfo {pages} {257} (\bibinfo {year} {2010})}\BibitemShut
  {NoStop}%
\bibitem [{\citenamefont {Balwada}\ \emph {et~al.}(2022)\citenamefont
  {Balwada}, \citenamefont {Xie}, \citenamefont {Marino},\ and\ \citenamefont
  {Feraco}}]{balwada2022direct}%
  \BibitemOpen
  \bibfield  {author} {\bibinfo {author} {\bibfnamefont {D.}~\bibnamefont
  {Balwada}}, \bibinfo {author} {\bibfnamefont {J.-H.}\ \bibnamefont {Xie}},
  \bibinfo {author} {\bibfnamefont {R.}~\bibnamefont {Marino}},\ and\ \bibinfo
  {author} {\bibfnamefont {F.}~\bibnamefont {Feraco}},\ }\bibfield  {title}
  {\bibinfo {title} {Direct observational evidence of an oceanic dual kinetic
  energy cascade and its seasonality},\ }\href@noop {} {\bibfield  {journal}
  {\bibinfo  {journal} {Science Advances}\ }\textbf {\bibinfo {volume} {8}},\
  \bibinfo {pages} {eabq2566} (\bibinfo {year} {2022})}\BibitemShut {NoStop}%
\bibitem [{\citenamefont {Dewar}\ and\ \citenamefont
  {Hogg}(2010)}]{dewar2010topographic}%
  \BibitemOpen
  \bibfield  {author} {\bibinfo {author} {\bibfnamefont {W.~K.}\ \bibnamefont
  {Dewar}}\ and\ \bibinfo {author} {\bibfnamefont {A.~M.}\ \bibnamefont
  {Hogg}},\ }\bibfield  {title} {\bibinfo {title} {Topographic inviscid
  dissipation of balanced flow},\ }\href@noop {} {\bibfield  {journal}
  {\bibinfo  {journal} {Ocean Modelling}\ }\textbf {\bibinfo {volume} {32}},\
  \bibinfo {pages} {1} (\bibinfo {year} {2010})}\BibitemShut {NoStop}%
\bibitem [{\citenamefont {Nikurashin}\ \emph {et~al.}(2013)\citenamefont
  {Nikurashin}, \citenamefont {Vallis},\ and\ \citenamefont
  {Adcroft}}]{nikurashin2013routes}%
  \BibitemOpen
  \bibfield  {author} {\bibinfo {author} {\bibfnamefont {M.}~\bibnamefont
  {Nikurashin}}, \bibinfo {author} {\bibfnamefont {G.~K.}\ \bibnamefont
  {Vallis}},\ and\ \bibinfo {author} {\bibfnamefont {A.}~\bibnamefont
  {Adcroft}},\ }\bibfield  {title} {\bibinfo {title} {Routes to energy
  dissipation for geostrophic flows in the southern ocean},\ }\href@noop {}
  {\bibfield  {journal} {\bibinfo  {journal} {Nature Geoscience}\ }\textbf
  {\bibinfo {volume} {6}},\ \bibinfo {pages} {48} (\bibinfo {year}
  {2013})}\BibitemShut {NoStop}%
\bibitem [{\citenamefont {Zhang}\ and\ \citenamefont
  {Xie}(2023)}]{zhang2023spectral}%
  \BibitemOpen
  \bibfield  {author} {\bibinfo {author} {\bibfnamefont {L.-F.}\ \bibnamefont
  {Zhang}}\ and\ \bibinfo {author} {\bibfnamefont {J.-H.}\ \bibnamefont
  {Xie}},\ }\bibfield  {title} {\bibinfo {title} {Spectral condensation in
  quasi-geostrophic turbulence above small-scale topography},\ }\href@noop {}
  {\bibfield  {journal} {\bibinfo  {journal} {arXiv preprint arXiv:2311.16612}\
  } (\bibinfo {year} {2023})}\BibitemShut {NoStop}%
\bibitem [{\citenamefont {{Renault}}\ \emph {et~al.}(2019)\citenamefont
  {{Renault}}, \citenamefont {{Marchesiello}}, \citenamefont {{Masson}},\ and\
  \citenamefont {{McWilliams}}}]{renault.marchesiello.ea_2019}%
  \BibitemOpen
  \bibfield  {author} {\bibinfo {author} {\bibfnamefont {L.}~\bibnamefont
  {{Renault}}}, \bibinfo {author} {\bibfnamefont {P.}~\bibnamefont
  {{Marchesiello}}}, \bibinfo {author} {\bibfnamefont {S.}~\bibnamefont
  {{Masson}}},\ and\ \bibinfo {author} {\bibfnamefont {J.~C.}\ \bibnamefont
  {{McWilliams}}},\ }\bibfield  {title} {\bibinfo {title} {Remarkable control
  of western boundary currents by eddy killing, a mechanical air-sea coupling
  process},\ }\href@noop {} {\bibfield  {journal} {\bibinfo  {journal}
  {Geophys. Res. Lett.}\ }\textbf {\bibinfo {volume} {46}},\ \bibinfo {pages}
  {2743} (\bibinfo {year} {2019})}\BibitemShut {NoStop}%
\bibitem [{\citenamefont {Cushman-Roisin}\ and\ \citenamefont
  {Beckers}(2011)}]{cushman2011introduction}%
  \BibitemOpen
  \bibfield  {author} {\bibinfo {author} {\bibfnamefont {B.}~\bibnamefont
  {Cushman-Roisin}}\ and\ \bibinfo {author} {\bibfnamefont {J.-M.}\
  \bibnamefont {Beckers}},\ }\href@noop {} {\emph {\bibinfo {title}
  {Introduction to geophysical fluid dynamics: physical and numerical
  aspects}}}\ (\bibinfo  {publisher} {Academic press},\ \bibinfo {year}
  {2011})\BibitemShut {NoStop}%
\end{thebibliography}%

\clearpage
\onecolumngrid

\section*{Supplementary Materials 1: Inertial Boundary Layers}

In the theory of inertial inflow boundary layers the relation between $\psi$ and $q$ is established in an intermediate connection area just outside the boundary layer, where we expect the same functional relation between $\psi$ and $q$ to hold as inside the boundary layer (for details see \cite{vallis2017atmospheric}). In the limit of vanishing boundary layer thickness, the dynamical balance in the boundary layer will be the advection of relative vorticity by the interior flow and the creation of relative vorticity by the $\beta$-term. The thickness then scales like $\delta_I \sim \sqrt{-U_{in}/\beta^*}$, and the outer matching yields $q = -\beta^*\psi/U_{in}$.\\

In the simulations we identify two areas in which a functional relation between $q$ and $\psi$ holds. To identify these regions, we draw the isolines $u, v = 0$ that cross the centre of the gyre (the maximum of $\langle \psi \rangle$ close to the western boundary). The inflow layer is defined as the area west of the line $v = 0$ and south of $u = 0$, whereas the outflow layer is defined as the area north of $u = 0$ and east of $v = 0$. We also cut out areas very close to the boundaries where we expect viscous effects to be dominant. The resulting areas are depicted in figure 3 in the main text. \\

For the outflow layer (with positive inflow velocity along the x-direction), a matching with the interior flow is not self-consistent as the boundary layer thickness becomes imaginary and the solution becomes oscillatory. Furthermore, the inertial region does not reach the western part of the domain. 

We noticed that this inertial region contains a jet meandering around a closed cyclonic recirculation zone. Calling $\delta_o$ the jet width,  calling  $U_{out}$ the jet velocity, and assuming continuity of the  mass transport between the western boundary layer and the meandering jet results in
\begin{equation}
\delta_o U_{out} = U_I\delta_I = U_{Sv}L.\label{eq:trans}
\end{equation}
We observe that the wavelength of jet meanders and the radius of the cyclonic recirculation zone are of the order of the jet width $\delta_o$.  Meanders can be interpreted as a doppler-shifted  Rossby wave, which is stationary when \begin{equation}\delta_o = \sqrt{U_{out}/\beta^*}.\label{eq:delo}\end{equation}
This length also corresponds to the typical size of a stationary vortex on a beta plane in the presence of a mean flow $U_{out}$. Combining Eq. (\ref{eq:delo}) and (\ref{eq:trans}) yields 
\begin{equation}
    \delta_o = \beta^{*-\frac{2}{3}}, U_{out} = \beta^{*-\frac{1}{3}} .
\end{equation}
To determine the functional relation between $\psi$ and $q$ in the outflow layer we consider the (non-inertial) region in the northwestern part of the domain. This region connects the western boundary layer in the inflow layer to the meandering jet of the outflow layer. Close to the outflow layer, we assume that the flow velocity is zonal with velocity $U_{out}>0$ and we neglect relative vorticity with respect to planetary vorticity (which is only marginally satisfied).  In that case we can apply the same reasoning as classical inertial layer theory with $\langle q\rangle=\beta^* y$  and $\langle \psi\rangle=-U_{out} (y-L)$, assuming $\psi=0$ at the northern wall  in the outflow layer. We hence retrieve the relation. 
\begin{equation}
   \langle q\rangle = -\beta^*/U_{out}\langle \psi\rangle  +\beta^*L .
\end{equation}
To test this relation, we diagnose the velocities $U_{in}$ and $U_{out}$ as the average zonal velocities along the isolines $v=0$ for each part of the gyres connected to the matching regions. This yields values of $U_{out} = 0.18$ and $U_{in} = 0.07$, for which the scaling theory gives the correct order of magnitude ($U_{out} \sim \beta^{*-\frac{1}{3}} = 0.21$, $U_{in} \sim \beta^{*-1} = 0.01$, prefactors of order $\pi$ not taken into account). Using the diagnosed values, the slopes given by inertial layer theory then agree well with the observed functional relations between $\langle\psi\rangle$ and $\langle q\rangle$ in both areas. The slight offset of the relation for the outflow may stem from the fact that the relative vorticity is not negligible in the matching region. \\

\section*{Supplementary Materials 2: Mean Flow Variations}

\begin{figure}[!htb]
\centering
\includegraphics[width = \textwidth]{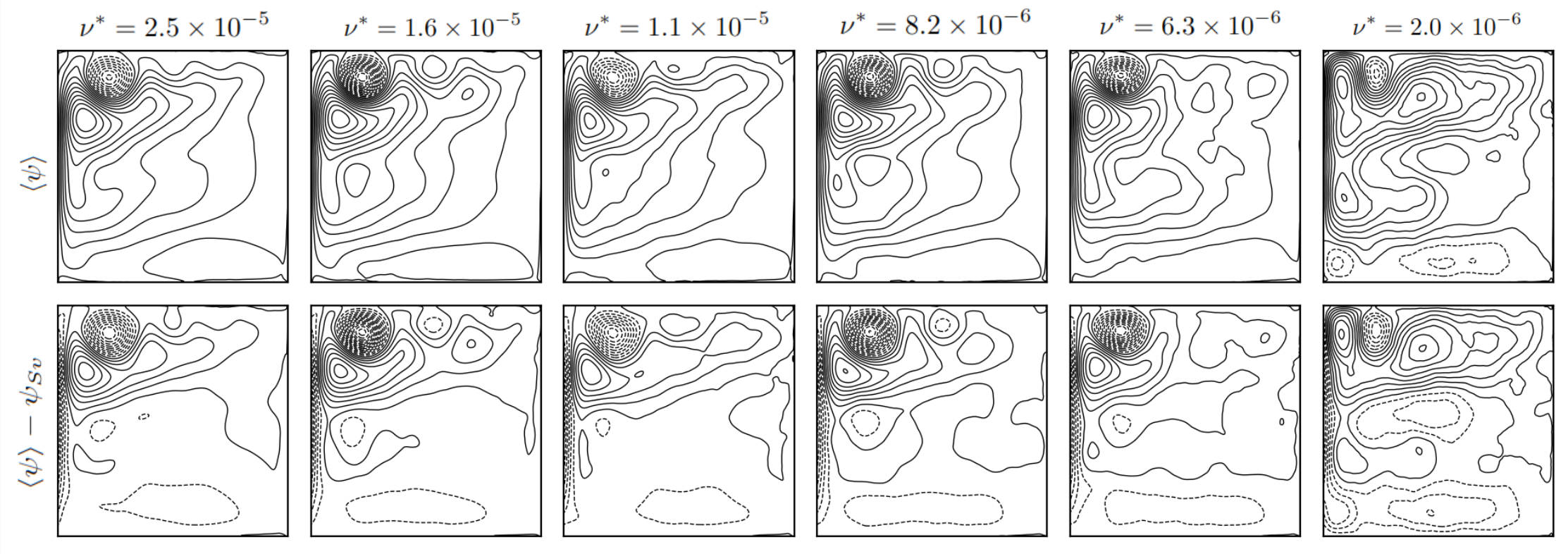}
\caption{Top line: Mean stream function $\langle \psi \rangle$ for different values of $\nu^*$. Bottom line: Deviations from Sverdrup Flow. Contour lines are plotted with intervals $\beta^{*-1}$.}
\label{fig: psi_mean}
\end{figure}

Although the energy of the mean flow remains constant its structure changes slightly when decreasing $\nu^*$ (figure \ref{fig: psi_mean}). In the interior, the linear response of the system to the forcing gives the Sverdrup solution of the form

\begin{equation}
    \psi_{Sv} = \frac{1}{\beta^*}\pi(1-x)\sin\left(\pi y\right)
\end{equation}

If we subtract this from the mean stream function $\langle \psi \rangle$ we observe that the departures from this solution are mainly zonal structures and hence suggest that they results from the non-linear interaction of Rossby basin modes as in \cite{fox2004wind2} (also no inertial relation between $\langle \psi \rangle$ and $\langle q \rangle$ was observed for these zonal structures). 

\section*{Supplementary Materials 3: Numerical Methods}

For the numerical implementation of the quasigeostrophic equations we apply standard finite-difference discretization procedure using Basilisk software (\url{http://basilisk.fr}). Vorticity and stream function are collocated at cell vertices and we use the Arakawa Jacobian for the advection term \cite{cushman2011introduction}. For the inversion of the elliptic equation we use a multigrid method. Time integration is performed with a second order predictor-corrector scheme. Impermeability conditions are achieved by imposing $\psi = 0$ at the edges when inversing $\omega$. The no-slip boundary condition is implemented by specifying the value of the vorticity on the domain boundaries for the viscous operator. By performing a Taylor expansion in the vicinity of the boundary (in this example, the western boundary) we get

\begin{equation}
  \label{eq:noslip}
  \psi_1 = \psi_0 + \Delta \left. \frac{\partial \psi}{\partial x}\right|_0 + \frac{\Delta^2}{2} \left. \frac{\partial^2 \psi}{\partial x^2}\right|_0 \, .
\end{equation}

So at this order, the vorticity at the edge is

\begin{equation}
  \omega_0 = \left. \frac{\partial^2 \psi}{\partial x^2}\right|_0 = 2 \frac{\psi_1}{\Delta^2}\, ,
  \label{edge}
\end{equation}
since $\psi_0 = 0$ (no flow) and $\left. \partial \psi / \partial x \right|_0 = 0$ (no slip). \\

In practice the methods outlined above work well in the domain interior, we therefore use them to timestep $\omega$ in the interior. We then inverse $\omega$ to obtain $\psi$, from which we can at last calculate the new values of $q$ at the boundaries with equation \ref{edge}.\\

Almost all simulations that are discussed in the main text were performed at a numerical resolution of $2048\times2048$ gridpoints. The only exception is the most turbulent run (at $\nu^* = 2\times 10^{-6}$), which was performed at a resolution of $8192 \times 8192$ gridpoints. To check for numerical convergence the simulation at $\nu^* = 6.25 \times 10^{-6}$ was relaunched with a doubled numerical resolution. Both its mean energy $\langle E\rangle$ and its mean dissipation $\langle \varepsilon \rangle$ changed by less than 5\%.

\section*{Supplementary Materials 4: Dissipative Area Fraction}

\begin{figure}[!htb]
\centering
\includegraphics[width = \textwidth]{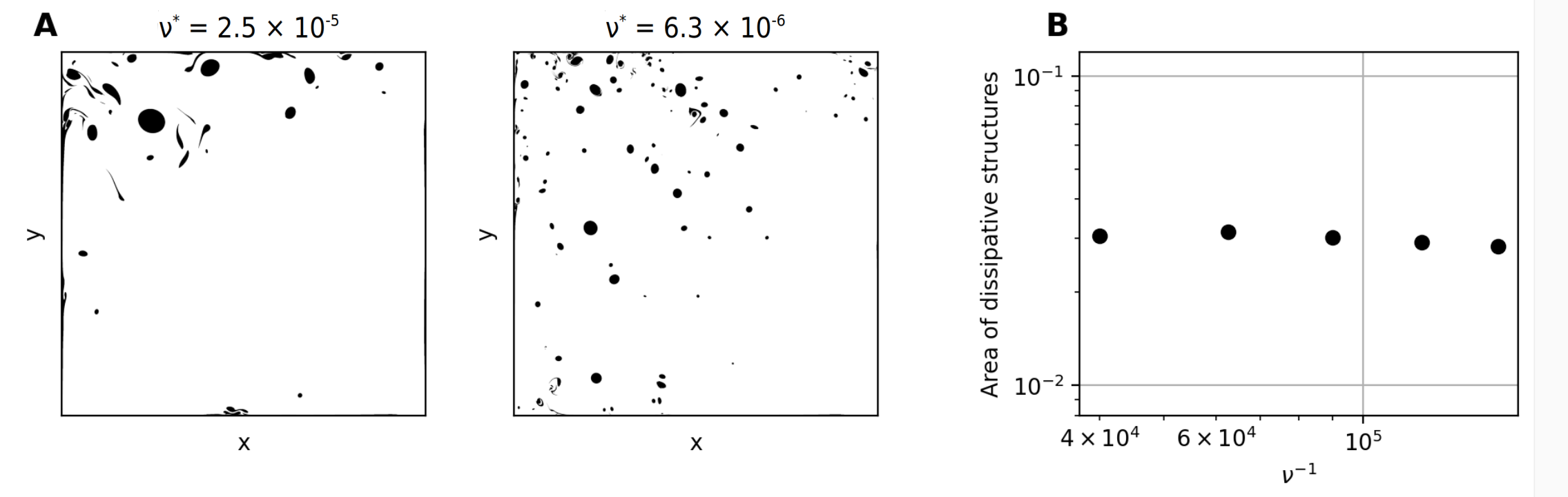}
\caption{\textbf{A}: Snapshots of dissipative structures at two values of $\nu^*$. To create the binary images, we compare vorticity values to $\omega_{max}/10$. \textbf{B} Area of dissipative structures against $\nu^*$. Note how, although the structures become smaller (as shown in A), the total area remains constant (the prediction of area fraction from scaling analysis is $\beta^{*-1} = 10^{-2}$).}
\label{fig: area_frac}
\end{figure}

The scaling analysis of the vortex gas presented in the main text is further validated by observing the area fraction of dissipative structures. We define the dissipative area fraction as the fractional area where $\omega > \omega_max/10$. A by-product of the scaling analysis is that the total vortex area $A$, which we diagnose with the dissipative area fraction, becomes independent of viscosity. It is given by 

\begin{equation}
    A \sim N\lambda^2 \sim \beta^{*-1},
\end{equation}

which we verify in figure \ref{fig: area_frac}.

\section*{Movie Legends}

\noindent\textbf{Movie S1.} Comparison of spin-up of flow regimes corresponding to free-slip and no-slip boundary conditions at equal values of $\nu^* = 10^{-5}$ and $\beta^* = 100$. On the left the free-slip simulation develops inertial runaway, with an energetic recirculation at the domain scale. The no-slip simulation exhibits Gyre Turbulence, with energetic small scales.\\

\noindent\textbf{Movie S2.} Animation of Gyre Turbulence in the Vortex Gas regime ($\nu^* = 2 \times 10^{-6}, \beta^* = 100$, Numerical Resolution: $8192 \times 8192$ grid points). Boundary layer detachments occur at all boundaries, but the northwestern corner remains the most active region, mostly due to the persistence of the strong cyclonic (red) vortex. With regards to the scaling theory of the vortex gas, notice the event at the start of the animation, where the cyclonic (red) vortex collides with the western boundary and detaches the viscous sublayer. 

\end{document}